\documentclass[prl,aps,twocolumn,superscriptaddress,notitlepage]{revtex4}
\usepackage{amsmath}
\usepackage{graphicx}
\usepackage{bm}
\usepackage{amsbsy}
\usepackage{amsfonts}
\usepackage{amsthm}
\usepackage[dvips]{color}

\newcommand{\ket}[1]{\left| #1\right\rangle}

\begin{document}

\title{Opto-mechanical micro-macro entanglement}

\author{R. Ghobadi}
\address{Institute for Quantum Information Science and
Department of Physics and Astronomy, University of Calgary,
Calgary T2N 1N4, Alberta, Canada}
\author{S. Kumar}
\address{Institute for Quantum Information Science and
Department of Physics and Astronomy, University of Calgary,
Calgary T2N 1N4, Alberta, Canada}
\author{B. Pepper}
\address{Department of Physics, University of California, Santa Barbara, California 93106, USA}
\author{D. Bouwmeester}
\address{Department of Physics, University of California, Santa Barbara, California 93106, USA}
\address{Huygens Laboratory, Leiden University, P.O. Box 9504, 2300 RA Leiden, Netherlands}
\author{A.I. Lvovsky}
\address{Institute for Quantum Information Science and
Department of Physics and Astronomy, University of Calgary,
Calgary T2N 1N4, Alberta, Canada}
\address{Russian Quantum Center, 100 Novaya St., Skolkovo, Moscow 143025, Russia}
\author{C. Simon}
\address{Institute for Quantum Information Science and
Department of Physics and Astronomy, University of Calgary,
Calgary T2N 1N4, Alberta, Canada}

\begin{abstract}
We propose to create and detect opto-mechanical entanglement by storing one component of an entangled state of light in a mechanical resonator and then retrieving it. Using micro-macro entanglement of light as recently demonstrated experimentally, one can then create opto-mechanical entangled states where the components of the superposition are macroscopically different. We apply this general approach to two-mode squeezed states where one mode has undergone a large displacement. Based on an analysis of the relevant experimental imperfections, the scheme appears feasible with current technology.
\end{abstract}
\maketitle

Vigorous efforts are currently being undertaken to bring quantum effects such as superposition and entanglement to the macroscopic level \cite{generalmacro,micromacro,lvovsky,bruno,demartini,RTS,sekatski,ghobadiPRL}. One prominent goal in this context is the creation of entanglement between a microscopic and a macroscopic system \cite{micromacro,lvovsky,bruno,demartini,RTS,sekatski,ghobadiPRL}, following Schr\"{o}dinger's famous thought experiment that involved a decaying nucleus and a cat \cite{schrodinger}.

In opto-mechanical systems the quantum regime has recently been reached \cite{verhagen,QOM,OMreview,chan}, but opto-mechanical entanglement has not yet been demonstrated. In a certain sense any entanglement of an opto-mechanical system can be seen as micro-macro entanglement, because the mechanical system always involves billions of atoms. However, for many proposals \cite{marshall,romero,pepper} the different components of the entangled state only differ by (of order) a single phonon.

Here we show how to create opto-mechanical micro-macro entanglement in a stronger sense by combining two key ideas. First, we propose a convenient method for both creating and detecting opto-mechanical entanglement, based on mapping one component of an entangled state of light onto the mechanical resonator and then retrieving it \cite{OMstorage,palomaki}. Demonstrating entanglement for the retrieved light then demonstrates the existence of opto-mechanical entanglement in the intermediate state. Second, we show that this approach makes it possible to create ``Schr\"{o}dinger cat'' type opto-mechanical entangled states where there is a macroscopic difference for a physical observable between the different components of the superposition, based on recent experiments demonstrating micro-macro entanglement of light \cite{lvovsky,bruno}. The physical observable in our case is the variance of the phonon number. Our proposal is thus different from Refs. \cite{cats-position}, which aim to create superposition states of mechanical systems with a large separation in position.

We propose to first create purely optical micro-macro entanglement by amplification of one component of an initial microscopic entangled state \cite{lvovsky,bruno,RTS,sekatski,demartini,ghobadiPRL}, and to then convert the photons in the amplified component into phonons. The entanglement can be verified by reconverting the phonons into photons and using the de-amplification and detection techniques of Refs. \cite{lvovsky,bruno}. De-amplification is advantageous in practice compared to trying to verify micro-macro entanglement by direct detection, which requires extremely high measurement precision \cite{raeisi}.

\begin{figure}
\centering{}\includegraphics[width=0.9\linewidth]{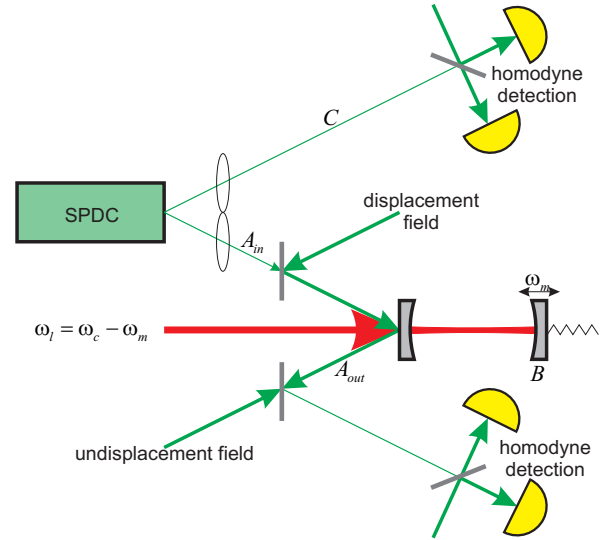}\caption{Proposed setup. The spontaneous parametric down-conversion source (SPDC) creates a (microscopic) two-mode squeezed state. One mode is directly detected by homodyne detection. The other mode is displaced by a macroscopic amount through the interference with a strong displacement field and then stored onto a mechanical oscillator using an opto-mechanical cavity and a strong red-detuned control beam. This creates opto-mechanical micro-macro entanglement. The state of the mechanical system can subsequently be reconverted into light. The entanglement is detected by first displacing the mode back to the microscopic level, followed by homodyne detection.
}
\label{setup}
\end{figure}

The general approach described above can be applied to different micro-macro entangled states \cite{lvovsky,bruno,RTS,sekatski,demartini,ghobadiPRL}. We illustrate it by introducing two-mode squeezed states where one mode has undergone a large displacement. We propose to use displacement as the amplification process because it creates states that are comparatively robust under photon loss \cite{lvovsky,bruno,sekatski}. Displaced two-mode squeezed states furthermore have the interesting property that the degree of entanglement and the degree of macroscopicity can be varied almost independently by choosing the amount of squeezing and the size of the displacement. Moreover these states are Gaussian, making it possible to quantify their entanglement exactly even in the presence of imperfections. See Appendix 1 for the application of our approach to displaced single-photon entanglement \cite{lvovsky,bruno,sekatski}.

We start by creating a two-mode squeezed state
$|\psi_0\rangle=\sqrt{1-t^{2}}\sum_{n=0}^{\infty}t^{n}|n\rangle_{A}|n\rangle_{C}$
with $t=\tanh(r)$ where $r$ is the squeezing strength. For moderate $r$ only the first few terms contribute significantly. For example, for $r=0.5$ (or 4.3dB of squeezing) the probabilities for the first terms are $p_{00}=0.786, p_{11}=0.168, p_{22}=0.036$, and the total weight of the remaining terms is only 0.001. We denote the optical modes $A$ and $C$, reserving the label $B$ for the mechanical oscillator. We apply the displacement operator $D(\alpha)=e^{\alpha a^{\dagger}-\alpha^* a}$ in mode $A$, generating the state
\begin{equation}
|\psi_D\rangle=\sqrt{1-t^{2}}\sum_{n=0}^{\infty}t^{n}(D(\alpha)|n\rangle)_{A}|n\rangle_{C}.
\label{dtms}
\end{equation}
The displacement can be implemented by interference with a strong coherent beam \cite{displacement,babichev,lvovsky,bruno}, see Fig. \ref{setup}. A displaced Fock state $D(\alpha)|n\rangle$ has a photon number variance $(2n+1)|\alpha|^2$. For moderate $r$ and large $\alpha$ the state (\ref{dtms}) is thus a superposition of a small number of relevant components that have macroscopically distinct photon number variances. Increasing the displacement $\alpha$ increases the macroscopicity of the superposition. On the other hand, increasing the squeezing parameter $r$ increases the entanglement of the state, in particular the number of components that contribute significantly. Here we focus on the moderate squeezing regime with only a few components, which is also the easiest regime to achieve experimentally.

The displaced mode $A$ of the two-mode squeezed state is now fed into a cavity and stored onto the mechanical mode $B$
using the opto-mechanical coupling between the cavity field and the mechanical mode \cite{OMstorage,palomaki}, see Fig. \ref{setup}.
The basic opto-mechanical Hamiltonian is
$H=\hbar \Delta a^{\dagger}a+\hbar\omega_{m}b^{\dagger}b+\hbar g_0 a^{\dagger}a(b+b^{\dagger})$,
where $\Delta=\omega_c-\omega_L$ is the detuning between the cavity resonance and the frequency of the control beam (see below), $a$ is the annihilation operator for the cavity mode, $\omega_m$ is the mechanical resonance frequency, $b$ is the mechanical mode annihilation operator, $g_0$ is the bare opto-mechanical coupling, and the Hamiltonian is written in the rotating frame with respect to the frequency of the control beam. If the control beam is red-detuned by $\omega_m$ with respect to the cavity resonance (and if $\omega_m \gg \kappa$, the resolved-sideband regime), one obtains the
effective beam splitter Hamiltonian
$H_{\mbox{eff}}=g (a^{\dagger}b+ab^{\dagger})$, where $g$ is proportional to $g_0$ and to the amplitude of the control beam \cite{OMreview,hofer}. The resulting equations of motion are
$\dot{a}=-\kappa a-igb+\sqrt{2\kappa}a_{in}$
and $\dot{b}=-iga$,
where we are ignoring mechanical damping and the related noise for now (see below). The input-output relation for the cavity is
$a_{out}=-a_{in}+\sqrt{2\kappa} a$. We consider the situation where the cavity decay rate $\kappa \gg g$ (and it is also much greater than the bandwidth of the input light). One can then adiabatically eliminate the cavity mode \cite{hofer,braunstein},
$a(t)=\frac{1}{\kappa}(-igb+\sqrt{2\kappa}a_{in})$.
This gives the equation of motion
$\dot{b}=-Gb-i\sqrt{2G}a_{in}$
with $G=g^2/\kappa$, and the input-output relation
$a_{out}=a_{in}-i\sqrt{2G}b$.
The solution is
$b(t)=-i\sqrt{2G}e^{-Gt}\int_0^{t} e^{Gt'}a_{in}(t')dt' + e^{-Gt} b(0)$ for the mechanical mode and $a_{out}(t)=-i\sqrt{2G}e^{-Gt} b(0)+a_{in}(t)-2G e^{-Gt} \int_0^t e^{Gt'}a_{in}(t')dt'$ for the output field.
Both storage and retrieval can be implemented by applying a constant coupling strength $G$ for a time duration $\tau$ (each).
It is convenient to introduce the modes
$A_{in}=\sqrt{\frac{2G}{e^{2G\tau}-1}}\int_0^{\tau} e^{Gt}a_{in}(t)dt$
and $A_{out}=\sqrt{\frac{2G}{1-e^{-2G\tau}}}\int_0^{\tau} e^{-Gt}a_{out}(t)dt$ \cite{hofer}, corresponding to the temporal modes in which the light should be prepared and detected respectively. We also introduce the notation $B_{in}=b(0)$ and $B_{out}=b(\tau)$. We then have the following equations for the storage (or write) process, $A_{out}^w=-i\sqrt{1-y^2} B_{in}^w+y A_{in}^w$, and
$B_{out}^w=y B_{in}^w-i\sqrt{1-y^2} A_{in}^w$,
where we have introduced the notation $y=e^{-G\tau}$ and the index $w$ indicates that these are the modes participating in the write process. Analogous equations (with index $r$) describe the read process. The two processes are linked by the identification $B_{in}^r=B_{out}^w$, which is exact for finite storage time in the absence of mechanical damping, or if no time passes between the write and read stages. This gives $A_{out}^r=-(1-y^2) A_{in}^w-i\sqrt{1-y^2}y B_{in}^w+y A_{in}^r$. Adopting the more compact notation $A_{out}=A_{out}^r$, $A_{in}=A_{in}^w-\alpha$ (so that $A_{in}$ refers to the optical input mode before the displacement), $B_{in}=B_{in}^w$ and $\delta A=A_{in}^r$ we finally have
\begin{equation}
A_{out}=-(1-y^2) (A_{in}+\alpha) - i y \sqrt{1-y^2} B_{in} + y \delta A.
\label{finalAout}
\end{equation}
Here $A_{out}$ is the optical output field after storage and retrieval (but before the eventual displacement back to the microscopic level), $A_{in}$ is the optical input, $B_{in}$ is the initial state of the mechanical oscillator, and $\delta A$ is an optical noise mode that is in the vacuum state.
One can see that the overall storage and retrieval efficiency is $(1-y^2)^2$ (in terms of photon number). This is very similar to the expressions obtained for the efficiency in other types of quantum memories \cite{qmemories}.

The displaced two-mode squeezed state is Gaussian. Its entanglement can therefore be quantified via the logarithmic negativity, which is defined as \cite{Adesso04}
$E_{N}=\mbox{max}\{0,-\ln(2\nu_{min})\}$,
where $\nu_{min}$ (the smallest symplectic eigenvalue of the partially
transposed covariance matrix) is given by $\nu_{min}=\sqrt{\frac{\Sigma-\sqrt{\Sigma^{2}-4detV}}{2}}$, where $\Sigma=detA+detB-2detC$, for the covariance matrix
$V=\left(\begin{array}{cc}
A & C\\
C^{T} & B\end{array}\right)$. The non-zero elements of the covariance matrix of the input state are determined by $\langle (X_A^{in})^2 \rangle=\langle (P_A^{in})^2 \rangle = \langle X_C^2 \rangle=\langle P_C^2 \rangle = \sinh^2(r)+\frac{1}{2}$, and $\langle X_A^{in} X_C \rangle =- \langle P_A^{in} P_C \rangle = \sinh(r) \cosh(r)$, where $X_A^{in}=(A_{in}+A_{in}^{\dagger})/\sqrt{2}$, $P_A^{out}=-i(A_{in}-A_{in}^{\dagger})/\sqrt{2}$, and $X_C, P_C$ are the quadrature operators for mode $C$. (Recall that $V_{11}=\langle (X_A^{in})^2 \rangle-(\langle X_A^{in}\rangle)^2$, $V_{13}=\langle X_A^{in} X_C \rangle-\langle X_A^{in}\rangle \langle X_C \rangle$ etc.)

Expressing Eq. (\ref{finalAout}) in terms of quadratures one has $X_A^{out}=-(1-y^2) (X_A^{in}+\sqrt{2}\alpha) + y \sqrt{1-y^2} P_B^{in} + y \delta X_A$ and $P_A^{out}=-(1-y^2) P_A^{in} - y \sqrt{1-y^2} X_B^{in} + y \delta P_A$, where $X_A^{out}, P_A^{out}, X_B, P_B, \delta X_A, \delta P_A$ are the quadrature operators corresponding to $A_{out}, B_{in}, \delta A$ respectively. It is then straightforward to determine the covariance matrix for the output modes and calculate the logarithmic negativity.
Note that in the above calculation the covariance matrix does not depend on the displacement, since $\alpha$ is fixed and mean values are subtracted in the definition of $V$. This changes however in the presence of phase noise, see below. Fig. \ref{OMcoupling} shows the entanglement in the final state as a function of $y$. One can see that there is a threshold for $y$ above which the entanglement becomes exactly zero. The value of this threshold depends on the initial phonon number $N_{in}=\langle \frac{(X_B^{in})^2+(P_B^{in})^2-1}{2} \rangle$ of the mechanical oscillator. Pre-cooling the mechanical oscillator close to the ground state is helpful for entanglement detection, but not strictly necessary. Note that the red-detuned control beam that is applied in the present protocol has a cooling effect \cite{chan}. Fig. \ref{OMcoupling} includes the effects of several other imperfections, namely phase noise, mechanical decoherence, in- and out-coupling loss, and loss on the micro side. We now discuss these effects in more detail.

 \begin{figure}
\centering{}\includegraphics[width=0.8\linewidth]{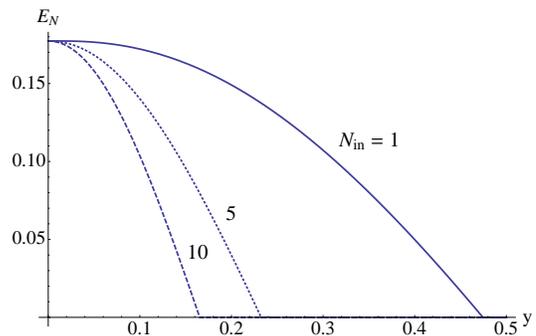}\caption{Entanglement
in the final state as a function of the opto-mechanical coupling parameter $y=e^{-G\tau}$, for different values of the initial mechanical phonon number $N_{in}$. In all cases $y$ has to be below a certain threshold value for entanglement to be observable, where the value of the threshold depends on $N_{in}$. The figure also includes the effect of other imperfections, the relevant parameter values are $x=\gamma/G=0.01$ and $N_{th}=10$ (mechanical noise), $\eta_1=\eta_2=\eta_c=0.8$ (losses), $\sigma=0.01$ (phase noise), see text for more discussion. The photon (or phonon) number corresponding to the displacement is $N_D=|\alpha|^2=5000$, and the squeezing parameter is $r=0.5$.}
\label{OMcoupling}
\end{figure}

\begin{figure}
\centering{}\includegraphics[width=0.8 \linewidth]{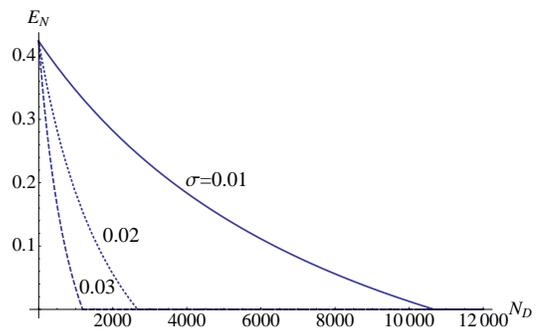}\caption{Entanglement in the final state as a function of the displaced photon (or phonon) number $N_D=|\alpha|^2$, for different values of the phase noise standard deviation $\sigma$. Phase noise limits the size of the displacement for which entanglement can be shown. Here $y=0.1$, $N_{in}=1$, and the other parameters are the same as in Figure 2.
}
\label{phasenoise}
\end{figure}

Phase noise can be modeled through the transformation $A_{out} \rightarrow e^{i\phi} A_{out}$, with a random phase $\phi$ with distribution $p(\phi)$. Let us assume that $p(\phi)$ is symmetric around $\phi=0$ and has a standard deviation $\sigma$, where $\sigma \ll 1$. The only term contributing to the covariance matrix that is significantly affected by the phase noise is $\langle (P_A^{out})^2 \rangle$, which gets an additional term $2 |\alpha|^2 (1-y^2)^2 \sigma^2$.  All other matrix elements only receive $O(\sigma^2)$ corrections (without the enhancement by the large $|\alpha|^2$ factor). However, this change in $\langle (P_A^{out})^2 \rangle$, which is not undone by the final displacement back to the microscopic level, has a significant effect on the entanglement, see Fig. \ref{phasenoise}. Phase noise limits the size of the displacement for which entanglement can be shown.
This increasing sensitivity to phase noise for increasing displacement provides further evidence (in addition to the above argument based on the photon/phonon number variances of the displaced Fock states) that the displaced two-mode squeezed state is indeed a macroscopic superposition state, see also Refs. \cite{lvovsky,bruno,sekatski,duer,sekatski2}. Ref. \cite{lvovsky} achieved very large displacements ($N_D > 10^8$) by using the same spatial mode (but orthogonal polarization modes) for the signal and displacement beam, leading to very high stability. This may be more challenging in the opto-mechanical context. In our examples we have picked $\sigma$ values more comparable to Ref. \cite{bruno}, where the signal and displacement beam were in separate spatial modes.

We will now take into account the mechanical damping and associated noise. The equation of motion for $b$ is now $\dot{b}=-\gamma b-iga+\sqrt{2\gamma}b_{in}$, leading to $\dot{b}=-G'b-i\sqrt{2G}a_{in}+\sqrt{2\gamma}b_{in}$ after adiabatic elimination of the cavity mode. Here $G'=G+\gamma$. Using techniques similar to Ref. \cite{hofer}, but keeping all orders of $\gamma$, one can show that
this equation together with the input-output relation for the cavity leads to the following modified equation for the optical output mode after storage and retrieval,
\begin{equation}
A_{out}=-\frac{1-y^2}{1+x} A_{in} - i \sqrt{\frac{1-y^2}{1+x}} y B_{in} + f_1 \delta A + f_2 \delta B,
\end{equation}
where we have introduced the notation $x=\gamma/G$, and $A_{in}$ is defined analogously to before, but using $G'$ instead of $G$. The modes $\delta A$ and $\delta B$ correspond to the optical and mechanical noise respectively, where the former is in the vacuum state, and the latter is in a thermal state at the temperature of the mechanical bath with a mean phonon number $N_{th}$. Their coefficients are $f_{1}=\frac{1}{1+x}\sqrt{x^{2}+y^{2}+\frac{4 xyG'\tau}{\sqrt{2\cosh(2G'\tau)-2}}}$ and
$f_{2}=\frac{1}{1+x}\sqrt{x(1+y^{2})+x(1-y^{2})^{2}-\frac{4xyG'\tau}{\sqrt{2\cosh(2G'\tau)-2}}}$. See Appendix 2 for more details on the calculation. Fig. \ref{heating} shows the effect of the mechanical noise on the entanglement in the final state; $x$ has to be below a certain threshold in order for entanglement to be present, where the value of the threshold depends on $N_{th}$. For the parameters of Fig. \ref{heating} one has the condition $N_{th}x=N_{th} \gamma/G \lesssim 0.2$; $N_{th}\gamma$ can be interpreted as the effective mechanical decoherence rate.

\begin{figure}
\centering{}\includegraphics[width=0.8\linewidth]{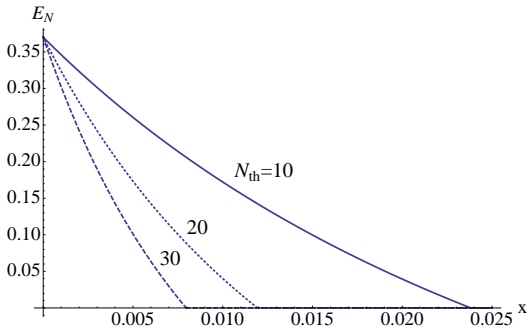}\caption{Entanglement in the final state as a function of the mechanical noise parameter $x=\gamma/G$, for different values of the bath mean phonon number $N_{th}$. The values of the other parameters are $N_D=|\alpha|^2=5000$, $r=0.5$, $y=0.1$, $N_{in}=1$, $\sigma=0.01$, and $\eta_1=\eta_2=\eta_c=0.8$.}
\label{heating}
\end{figure}

\begin{figure}
\centering{}\includegraphics[width=0.8\linewidth]{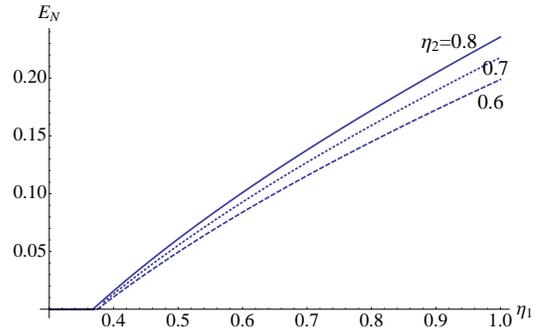}\caption{Entanglement in the final state as a function of $\eta_1$ for different values of $\eta_2$. Here $1-\eta_1$ and $1-\eta_2$ are the photon loss before and after the opto-mechanical system respectively. The values of the other parameters are $N_D=|\alpha|^2=5000$, $r=0.5$, $y=0.1$, $N_{in}=1$, $\sigma=0.01$, $x=0.01$, $N_{th}=10$, and $\eta_c=0.8$.
}
\label{loss}
\end{figure}

Another important imperfection is photon loss, including coupling losses for the opto-mechanical cavity and detection inefficiency. We model these losses by three beam splitters, one before and one after the opto-mechanical system in mode $A$ (with transmissions $\eta_1$ and $\eta_2$ respectively), as well as one in mode $C$. It is straightforward to include these effects in the covariance matrix of the final state. Fig. \ref{loss} shows that $\eta_1$ has to be above a certain threshold value (of order 0.4 for our choice of parameters) in order to be able to demonstrate entanglement. There is no equivalent condition for $\eta_2$ or $\eta_c$, but they also reduce the entanglement.

We propose an implementation based on the integrated optical and mechanical nanoscale resonator of Ref. \cite{chan} and the narrowband cavity-enhanced parametric down-conversion source of Ref. \cite{riedmatten}. Taking $\omega_m=2\pi\times3.7$ GHz, $\kappa=2\pi\times 500$ MHz, and $\gamma=2\pi \times 35$ kHz from Ref. \cite{chan} and assuming a bath temperature $T=2$ K, which is accessible with fairly simple cryostats, one has $N_{th} \approx 10$. The highest drive power used in Ref. \cite{chan} corresponds to $g \approx 2\pi\times 40$ MHz, leading to an effective coupling $G=g^2/\kappa \approx 2\pi \times 3.2$ MHz.
This gives $x=\gamma/G \approx 0.01$.  We propose $\tau \approx 100$ ns, which is in good correspondence with the PDC source of Ref. \cite{riedmatten}. This gives $y=e^{-G\tau}\approx 0.1$. Concerning photon loss,
Ref. \cite{cohen} already demonstrated of order 75\% in-coupling efficiency and 52 \% out-coupling efficiency in a system very similar to that of Ref. \cite{chan}, and even higher values for the coupling efficiencies should be possible \cite{cohen}.
We have neglected the effects of the squeezing part of the opto-mechanical Hamiltonian. They are expected to be suppressed by a factor
 $(\frac{\kappa}{\omega_m})^2$, which is less than 0.02 for the system parameters given above, justifying the approximation for this proposed implementation. Beam-splitter type opto-mechanical coupling was also demonstrated e.g. in Refs. \cite{verhagen,palomaki}. The creation and detection of opto-mechanical micro-macro entanglement is thus within reach of current technology.

The approach based on opto-mechanical storage and retrieval also allows one to conceive experiments that would test proposals for quantum gravity induced wave function collapse. For example, using the approach of Ref. \cite{trampolines} it is realistic to fabricate trampoline resonators with an effective mass of 500 ng, a mechanical frequency $\omega_m=2\pi\times 10$ kHz and a mechanical quality factor of $10^6$. At a temperature of 1 mK the environmentally induced decoherence timescale $1/N_{th}\gamma$ of 7.6 ms is then significantly longer than the decoherence times predicted for this system by the quantum gravity induced collapse models of Ref. \cite{ellis} (240 $\mu$s) and of Ref. \cite{penrose} (95 $\mu$s), see also Refs. \cite{collapse}. The latter number is obtained using the nuclear radius to define the mass distribution following Ref. \cite{diosi}. For a cavity length of 10 cm, a cavity finesse of $10^6$ and a control field power of 40 pW one can then have $\kappa \approx 2\pi\times 1.5$ kHz  and $G \approx 2\pi\times 200$ Hz, satisfying $\omega_m \gg \kappa \gg G \gg \gamma N_{th}\approx 2\pi \times 20$ Hz, as required for sideband cooling, adiabatic elimination of the cavity, and entanglement detection respectively. These parameters require a source of sub-kHz bandwidth two-mode squeezed light, which should be feasible based on parametric down-conversion with a narrowband pump laser in combination with filter cavities. Compared to the proposal of Ref. \cite{pepper}, which may also allow testing collapse models with weakly coupled opto-mechanical systems, the present approach has the advantage of not requiring any post-selection.

While the above-mentioned collapse times are not sensitive to the size of the displacement $\alpha$, varying $\alpha$ and hence the number of phonons involved in the superposition would also allow one to look for other types of deviations from quantum physics that might manifest in the little explored regime of superpositions of macroscopically different quantum numbers.

This work was supported by AITF, NSERC, NSF grant PHY-1206118, and NWO VICI grant 680-47-604. We thank P. Barclay for useful discussions.

{\it Notes added.} During the completion of this work, we became aware of related work by P. Sekatski and co-workers. Our two papers were jointly submitted to Physical Review Letters on August 30, 2013. 

After this work was completed we learned about the recent experiment of Ref. \cite{lehnertent}, where opto-mechanical entanglement in the microwave domain (created via blue-detuned driving, not by storing an entangled signal) is also detected by mapping the state of the mechanical mode onto the microwave field.

\section{Appendix 1: Displaced single-photon entanglement}

The procedure of interconverting the macroscopic part of a micro-macro entangled state between an optical and a mechanical system can be extended to a wide range of states. As an example, consider a displaced single-photon entangled state as created in the recent experiments of Refs. \cite{lvovsky,bruno},
\begin{equation}\label{DispNLP}
\ket{\psi_D}=\frac 1{\sqrt 2}[D(\alpha)\ket 1]_A\ket 0_C+[D(\alpha)\ket 0]_A\ket 1_C,
\end{equation}
which can be prepared by entangling the single-photon Fock state and the vacuum state on a beam splitter and subsequently displacing one of the output modes. This is a macroscopic superposition state for the same reasons as discussed for the displaced two-mode squeezed state in the paper, but with two components that have the same weight. Similarly to the two-mode squeezed state, mode $A$ of state (\ref{DispNLP}) can be stored in a mechanical vibration mode by means of the procedure described in the paper, resulting in a macroscopic optomechanical entangled state. Its entanglement can be verified by converting the vibration back to the optical domain and reversing the phase-space displacement.

\begin{figure}[t]
\includegraphics[width=\columnwidth]{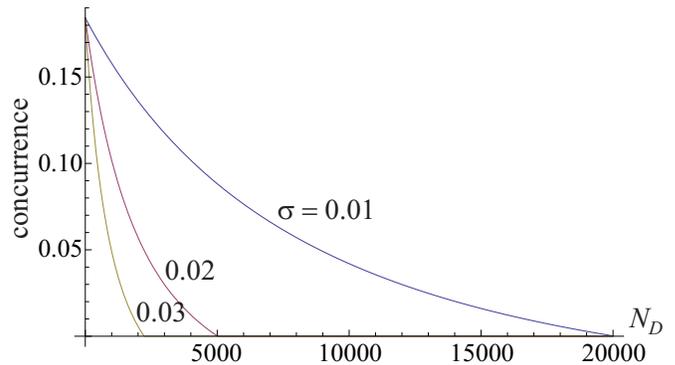}
\caption{\label{NLPFig}Entanglement of the truncated final state for the case of displaced single-photon entanglement as a function of the number of quanta $N_D=|\alpha|^2$ in the initial displacement, for different magnitudes $\sigma$  of the phase noise. The experimental parameters are identical to those used in Fig.~3 of the paper: $x=0.01$, $y=0.1$, $N_{th}=10$, $N_{in}=1$, $\eta_1=\eta_2=0.8$.}
\end{figure}

In order to study the behavior of entanglement in the resulting state with respect to various experimental parameters, we calculated its density matrix in the Fock basis using Eq.~(3) in the paper. Given the Fock basis decompositions of the input optical state, the thermal states in modes $B_{in}$ and $\delta B$, as well as the vacuum state in mode  $\delta A$, the state of mode $A_{out}$ is readily determined using the Fock representation of the beam splitter operator \cite{LeonhardtBook}.

Because the magnitude of the ``undisplacement" after the retrieval is chosen to cancel the effect of the initial displacement, the presence of these operations can be ignored in the calculation, aside from the effect of phase noise. Phase noise manifests itself in a random phase-space displacement along the momentum axis, with the displacement magnitude $\Delta P_A^{out}$ following a Gaussian distribution with variance $2|\alpha|^2(1-y^2)^2\sigma^2$. The cancelation of the macroscopic displacements permits us to perform the calculation in a subspace of the optical Hilbert space truncated to $n_{max}=15$ photons.

The entanglement of the final state can be determined by projecting that state onto the qubit subspace spanned by the vacuum and single-photon states, and evaluating the concurrence of the resulting two-qubit entangled state \cite{ghobadiPRL}. This concurrence exhibits a behavior that is quite similar to that of the logarithmic negativity of the displaced two-mode squeezed state studied in the paper. For example, as shown in Fig.~\ref{NLPFig}, it is sensitive to phase noise; the higher the noise amplitude, the lower the threshold displacement beyond which the entanglement disappears.

\section{Appendix 2: Mechanical Noise Calculations}

We here provide more details on the calculations leading to Eq. (3) in the paper, which takes into account the effects of mechanical decoherence. As explained in the paper, the basic equations for the opto-mechanical interaction in this case (after adiabatic elimination of the cavity mode) are
\begin{equation}
\dot{b}=-G^{'}b-i\sqrt{2G}a_{in}+\sqrt{2\gamma}b_{in}
\label{bdot}
\end{equation}
with $G'=G+\gamma$, and
\begin{equation}
a_{out}(t)=a_{in}(t)-i\sqrt{2G}b(t)
\label{output}
\end{equation}
The solution to Eq. (\ref{bdot}) is given by
\begin{equation}
b(t)=e^{-G^{'}t}\{b(0)+\intop_{0}^{t}dt' e^{G^{'}t'}[\sqrt{2\gamma}b_{in}(t^{'})-i\sqrt{2G}a_{in}(t^{'})]\}
\label{bt}
\end{equation}

It is convenient to describe the {\it write} process in terms of the following time-independent mode annihilation operators \cite{hofer}:

\begin{equation}
A_{out}^{(-)w}=\sqrt{\frac{2G^{'}}{1-e^{-2G^{'}\tau}}}\intop_{0}^{\tau}dte^{-G^{'}t}a_{out}(t)
\label{Caw-}
\end{equation}

\begin{equation}
A_{in}^{(+)w}=\sqrt{\frac{2G^{'}}{e^{2G^{'}\tau}-1}}\intop_{0}^{\tau}dte^{G^{'}t}a_{in}(t)
\end{equation}

\begin{equation}
 A_{in}^{(-)w}=\sqrt{\frac{2G^{'}}{1-e^{-2G^{'}\tau}}}\intop_{0}^{\tau}dte^{-G^{'}t}a_{in}(t)
\end{equation}

\begin{equation}
\delta B_{in}^{(-)w}=\sqrt{\frac{2G^{'}}{1-e^{-2G^{'}\tau}}}\intop_{0}^{\tau}dte^{-G^{'}t}b_{in}(t)
\end{equation}

\begin{equation}
\delta B_{in}^{(+)w}=\sqrt{\frac{2G^{'}}{e^{2G^{'}\tau}-1}}\intop_{0}^{\tau}dte^{G^{'}t}b_{in}(t)
\label{CBin+}
\end{equation}

In term of these new modes and using Eqs.(\ref{output},\ref{bt}), we obtain

\begin{eqnarray}
&&A_{out}^{(-)w}=(1-\frac{G}{G^{'}})A_{in}^{(-)w}-i\sqrt{\frac{G}{G^{'}}(1-y^{2})}B_{in}^{w}\nonumber\\
&&+\frac{G}{G^{'}}yA_{in}^{(+)w}-i\sqrt{\frac{\gamma G}{G^{'2}}}(y\delta B_{in}^{(+)w}-\delta B_{in}^{(-)w})
\label{Caoutw}
\end{eqnarray}

and

\begin{equation}
B_{out}^{w}=yB_{in}^{w}+\sqrt{\frac{\gamma}{G^{'}}(1-y^{2})}\delta B_{in}^{(+)w}-i\sqrt{\frac{G}{G^{'}}(1-y^{2})}A_{in}^{(+)w}
\label{boutw}
\end{equation}

with $B_{in}^w=b(0)$ and $B_{out}^w=b(\tau)$.

Note that if we neglect the mechanical damping in Eq. (\ref{Caoutw}), the first and last terms on the right-hand side become zero and Eq.(\ref{Caoutw}) becomes identical to the equation given for $A_{out}^w$ in the main text, with the idenfication $A_{out}^{(-)w}=A_{out}^w$, $A_{in}^{(+)w}=A_{in}^w$. We also note that Eqs.(\ref{Caoutw},\ref{boutw}) satisfy the canonical bosonic commutation relations, $[A_{out}^{(-)w},A_{out}^{(-)w\dagger}]=[B_{out}^{w},B_{out}^{w\dagger}]=1$.

The {\it read} process is formally identical to the write process, but with a different initial state for the mechanical oscillator and the optical input in the vacuum state. Identifying $B_{in}^{r}=B_{out}^{w}$ and using Eqs.(\ref{Caoutw},\ref{boutw}), we obtain

\begin{eqnarray}
&&A_{out}^{(-)r}=(1-\frac{G}{G^{'}})\delta A_{in}^{(-)r}-i\sqrt{\frac{G}{G^{'}}(1-y^{2})}B_{out}^{w}\nonumber\\&&+\frac{G}{G^{'}}y\delta A_{in}^{(+)r}-i\sqrt{\frac{\gamma G}{G^{'2}}}(y\delta B_{in}^{(+)r}-\delta B_{in}^{(-)r})
\label{Caoutr}
\end{eqnarray}

where $\delta A_{in}^{(-)r}$, $\delta A_{in}^{(+)r}$, $\delta B_{in}^{(+)r}$, $\delta B_{in}^{(-)r}$  are defined in analogy with Eqs.(\ref{Caw-}-\ref{CBin+}) and correspond to optical vacuum noise and mechanical thermal noise during the read-out process. Adopting the notation $A_{out}=A_{out}^{(-)r}$, $B_{in}=B_{in}^{w}$ and using Eq. (\ref{boutw}) we obtain

\begin{equation}
A_{out}=-\frac{1-y^{2}}{1+x}A_{in}-i\sqrt{\frac{1-y^{2}}{1+x}}yB_{in}+f_{1}\delta A+f_{2}\delta B
\label{aoutr}
\end{equation}

where
\begin{equation}
\delta A=\frac{1}{f_{1}(1+x)}(x\delta A_{in}^{(-)r}+y\delta A_{in}^{(+)r})
\end{equation}

\begin{equation}
\delta B=-\frac{i\sqrt{x}}{f_{2}(1+x)}(y\delta B_{in}^{(+)r}-\delta B_{in}^{(-)r}+(1-y^{2})\delta B_{in}^{(+)w})
\end{equation}

The coefficients $f_{1}$ and $f_{2}$ (given in the main text) are determined by demanding the canonical comutation relations $[\delta A,\delta A^{\dagger}]=[\delta B,\delta B^{\dagger}]=1$.

\end{document}